\documentclass{elsart3}

\usepackage{natbib}
\usepackage{psfig}
\usepackage{graphicx}

\usepackage{amssymb}
\newcommand{\aap}{A\&A\ }
\newcommand{\aaps}{A\&AS\ }

\newcommand{\apj}{ApJ\ }
\newcommand{\apjl}{ApJ\ }

\newcommand{\pasj}{PASJ\ }

\begin{document}

\begin{frontmatter}



\title{Detection of thermal X-ray emission in the halo of the plerionic supernova remnant G21.5-0.9}


\author[label2]{F. Bocchino}

\address[label2]{INAF - Osservatorio Astronomico di Palermo, Piazza del Parlamento 1, 90134 Palermo, Italy
}
\begin{abstract}
The detection of a soft thermal X-ray component in the spectrum of a
bright knot in the halo of the plerion G21.5-0.9 is reported. Using a
collisional ionization equilibrium model for an hot optically thin plasma,
a temperature $kT=0.12-0.24$ kev, a mass of 0.3--1.0 M$_\odot$ and a
density of 1.6-6 cm$^{-3}$ is derived. The spectral analysis suggests
a possible overabundance of Silicon with respect to the solar value
in the knot; if this will be confirmed this object may be a clump of
shocked ejecta.

\end{abstract}

\begin{keyword}
PWN \sep ejecta \sep G21.5-0.9 \sep plerion \sep X-rays

\end{keyword}

\end{frontmatter}


\section{Introduction and data analysis}

The plerion G21.5-0.9 has been the subject of various studies in radio
(\citealt{bwd01} and references therein), infrared (\citealt{gt98})
and X-ray (\citealt{shp01,lm02}). However, its weak X-ray halo was
discovered only recently (\citealt{scs00,wbb01}) thanks to the larger
sensitivity of Chandra (\citealt{wtv00}) and XMM-Newton (\citealt{jla01})
with respect to previous X-ray astronomy satellites. The weakness of the
halo has prevented a detailed study of its features up to now.  Luckily,
G21.5-0.9 is a calibration source for both Chandra and XMM-Newton and a
large number of calibration datasets have been accumulated and is still being
accumulated to this purpose.  In this paper, this large collection
of public data is used to investigate the nature of the brightest spot in the
X-ray halo of this plerion, namely the North Spur (\citealt{wbb01}).

For XMM-Newton, the data consist of the observation of G21.5-0.9
with the source positioned on the axis and off the axis of the EPIC
cameras (observation ID 0122700101, 0122700201, 0122700301, 0122700401,
0122700501), which have a total summed live time of 142 ks for PN and
146 ks for the MOS cameras. The observations were taken on April 7--17,
2000. The original event files were screened to eliminate the contribution
of soft proton flares. After the screening, the summed exposure time
were 79 ks for PN and 116 ks for MOS. All the analysis of XMM data
was performed with the software SAS v6.0.  The XMM-Newton spectra of
the North Spur, collected in the single observations using a circular
region of 20 arcsec radius centered at location 18$^h$33$^m$32.7$^s$
RA and -10$^d$32$^m$49$^s$ DEC (J2000), were summed and their response
averaged. The spectra of the two MOS cameras were also summed and
their effective areas and response matrices were averaged.  The fit
was performed separately for the PN and MOS, and then, after checking
that they give consistent results, they were fitted simultaneously.
The spectra have been rebinned to a minimum of 20 counts per bin and
the 0.5--9 keV energy range was selected for the spectral fitting.
The background used for the spectral analysis was collected in an annulus
centered on $18^h33^m33.8^s$ RA and $-10^d34^m07^s$ DEC J2000 between
85$^{\prime\prime}$ and 120$^{\prime\prime}$ from the center (this region
is inside the X-ray halo at the same average off-axis angle as the North
Spur). This method subtracts the contribution of the underlying continuum
due to diffuse emission of the X-ray halo from the source spectrum,
and concentrate the study on the intrinsic spectral properties of the
North Spur.

For Chandra, 21 observations from the Chandra archive for
which the source was located on the ACIS S3 chip were selected, in order to maximize
the instrument response in the soft X-ray band, and at off-axis angle
$<5^\prime$. The sequence number of the selected observation are 0159,
1433,  1554,  1717,  1769,  1771, 1839,  2873, 3693, 4353, 5166,1230,
1553, 1716, 1718, 1770, 1838, 1840, 3474,3700, 4354. The datasets
include the dataset used by \citet{shp01}, the latest report on X-ray
data of the North Spur. However, they used only 6 observation for a total
of 65 ks, whereas we also include all the more recent observation that
have been done, for a total of 196.5 ks. This is the reason why they did
not detect the thermal emission reported in this work. The observations
were reprocessed using CIAO 3.1 and CALDB 2.28 in order to apply the
most recent calibrations to all the data, including time dependent gain
correction. Afterwards, the data were filtered to retain grade 0,2,3,4,6
and the status column equal zero.  All the filtered datasets were merged
together using the CIAO {\sc merge\_all} task. The spectrum of the North
Spur was extracted using a circular region centered at the same location
as before, but with a smaller radius (13 arcsec) to exploit the high
resolution of the Chandra mirrors. The spectrum was rebinned like the
EPIC spectrum and the band 0.5--8 keV was used. The response matrix and
effective area were generated using the CIAO tasks for the analysis of
extended sources. The background was selected in the same way as the
XMM-Newton background.  All the fits were performed using XSPEC v11.3.

\section{Results of spectral fitting}

The spectra were first fitted with a power law model modified by
the interstellar absorption of \citet{mm83} and a reasonable fit was
found. All the best-fit parameters and their uncertainties are reported
in Table \ref{res}.  However, given the non uniform distribution of
residuals in the soft band and the possibility that the North Spur may
show signs of thermal emission due to hot plasma, as part of a shell
or ejecta associated to G21.5-0.9, the spectra were fitted with a
combination of a power law model and the {\sc mekal} optically-thin
thermal plasma model (\citealt{mgv85}, \citealt{log95}). The results,
also showed in Table \ref{res}, indicate that the two components fit is
better then the power law only fit. The decrement of $\chi^2$ is
significant according to the F-test (probability $\sim 10^{-4}$),
and the distribution of residuals is more homogeneous. The discrepancy
between XMM-Newton and Chandra derived fluxes are due to different
extraction region area and suggests that the source is extended
(especially its non-thermal emission).

\begin{table*}
  \caption{Summary of spectral fitting results to the X/ray spectrum of the bright knot ``North Spur.}
\label{res}
\medskip
\centering\begin{minipage}{12.2cm}
\begin{tabular}{lcccccc} \hline
\small
  Model & $N_H$  & $\gamma$ & Norm & $kT$ & flux\footnote{Absorbed flux due to the thermal component only, in the 0.5--2.0 keV.} & $\chi^2/dof$ \\
      & cm$^{-2}$ & photon & cm$^{-2}$ s$^{-1}$ keV$^{-1}$ & keV & erg cm$^{-2}$ s$^{-1}$  \\
      & $\times 10^{22}$ & index & $\times 10^{-4}$  & & $\times 10^{-15}$ & \\ \hline

\multicolumn{7}{c}{XMM-Newton PN and MOS} \\

Power-law & $1.45\pm0.13$ & $1.91\pm0.12$ & $13.9\pm2.5$ & - & - & 442/407 \\
P.L.+MEKAL & $2.63\pm0.30$ & $2.30\pm0.15$ & $29.3\pm6.5$ & $0.15\pm0.03$ & $12.0\pm4.5 $ & 423/405 \\
P.L.+VMEKAL\footnote{Only Mg and Si abundances were fitted (best-fit values and 90\% uncertainties are 1.0(0.1--2.3) and 4(1.3--8), respectively). Flux of Mg XI line at 1.34 keV is $3.4^{+2.0}_{-1.8}\times 10^{-6}$ ph cm$^{-2}$ s$^{-1}$, whereas flux of Si XIII line at 1.86 keV is $8.7^{+6.2}_{-5.3}\times 10^{-6}$ ph cm$^{-2}$ s$^{-1}$.} & $2.63\pm0.39$ & $2.26\pm0.13$ & $28.5\pm7.6$ & $0.17\pm0.03$ & $11.4\pm5.6$ & 410/403 \\

\multicolumn{7}{c}{Chandra ACIS S3} \\

Power-law & $1.15\pm0.09$ & $1.72\pm0.10$ & $9.3\pm1.2$ & - & - & 237/187 \\
P.L.+MEKAL & $1.78\pm0.32$ & $2.04\pm0.16$ & $15.4\pm3.7$ & $0.17\pm0.04$ & $5.3\pm1.3$ & 217/185 \\
P.L.+VMEKAL\footnote{Only Mg and Si abundances were fitted (90\% confidence range is $>0.6$ for Mg and $>5$ for Si). Flux of Mg XI line at 1.34 keV is $2.8^{+1.5}_{-2.5}\times 10^{-6}$ ph cm$^{-2}$ s$^{-1}$, whereas flux of Si XIII line at 1.86 keV is $9.5^{+3.7}_{-4.2}\times 10^{-6}$ ph cm$^{-2}$ s$^{-1}$.} & $1.81\pm0.24$ & $2.01\pm0.14$ & $14.8\pm3.0$ & $0.15\pm0.03$ & $7.9\pm2.3$ & 206/183 \\

\hline

\end{tabular}
\end{minipage}
\end{table*}

   \begin{figure}[bp!]
   \centerline{
      \includegraphics[width=6.5cm,angle=-90]{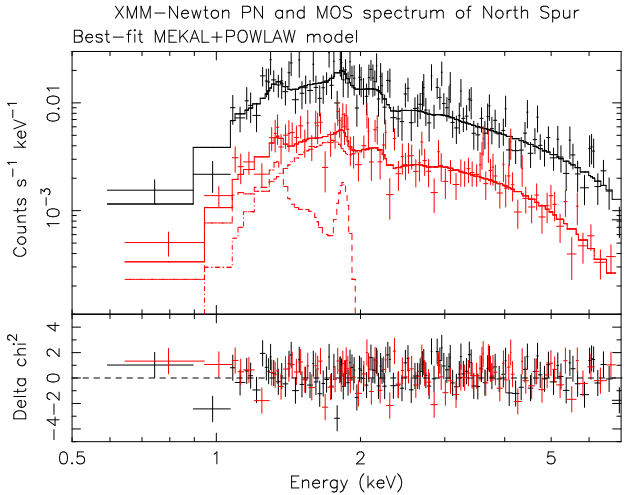}
      }
   \caption{XMM-Newton PN and MOS spectra of the bright knot named North
   Spur in the X-ray halo of the plerion G21.5-0.9. Solid line corresponds
   to the best-fit obtained with a thermal+non thermal model. We also
   show the contribution of thermal and non-thermal components to the
   total best-fit spectra (for MOS only).}

   \label{epic}
   \end{figure}

   \begin{figure}[bp!]
   \centerline{
      \includegraphics[width=9.5cm,angle=-90]{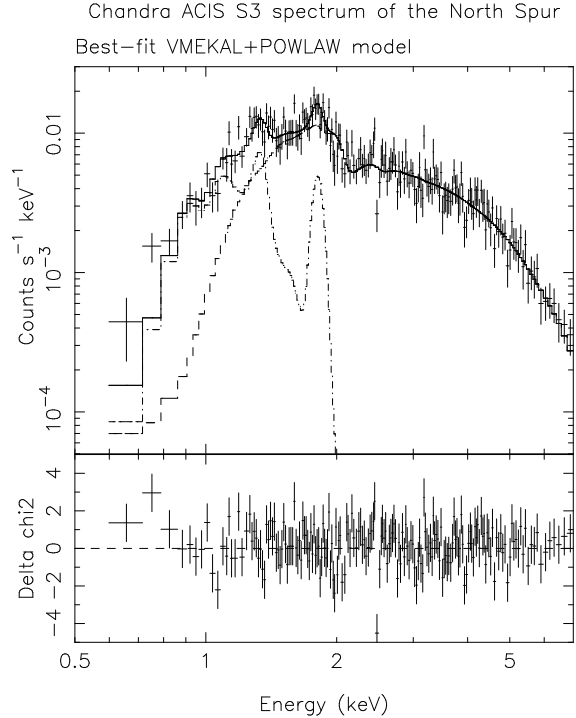}
      }
   \caption{Same as Fig. \protect\ref{epic} but for Chandra ACIS S3.}
   \label{acis}
   \end{figure}

The XMM-Newton and Chandra spectra are shown in Fig. \ref{epic} and
\ref{acis}, along with the best-fit two component model.  Emission
lines of Mg XI at 1.34 keV and Si XIII at 1.86 keV may be identified in
FIg. \ref{acis}. Other expected emission lines at different energies are
not evident due to limited statistics of these highly absorbed spectra
and moderate CCD spectral resolution.

   \begin{figure}[tbp!]
   \centerline{\includegraphics[width=8.0cm]{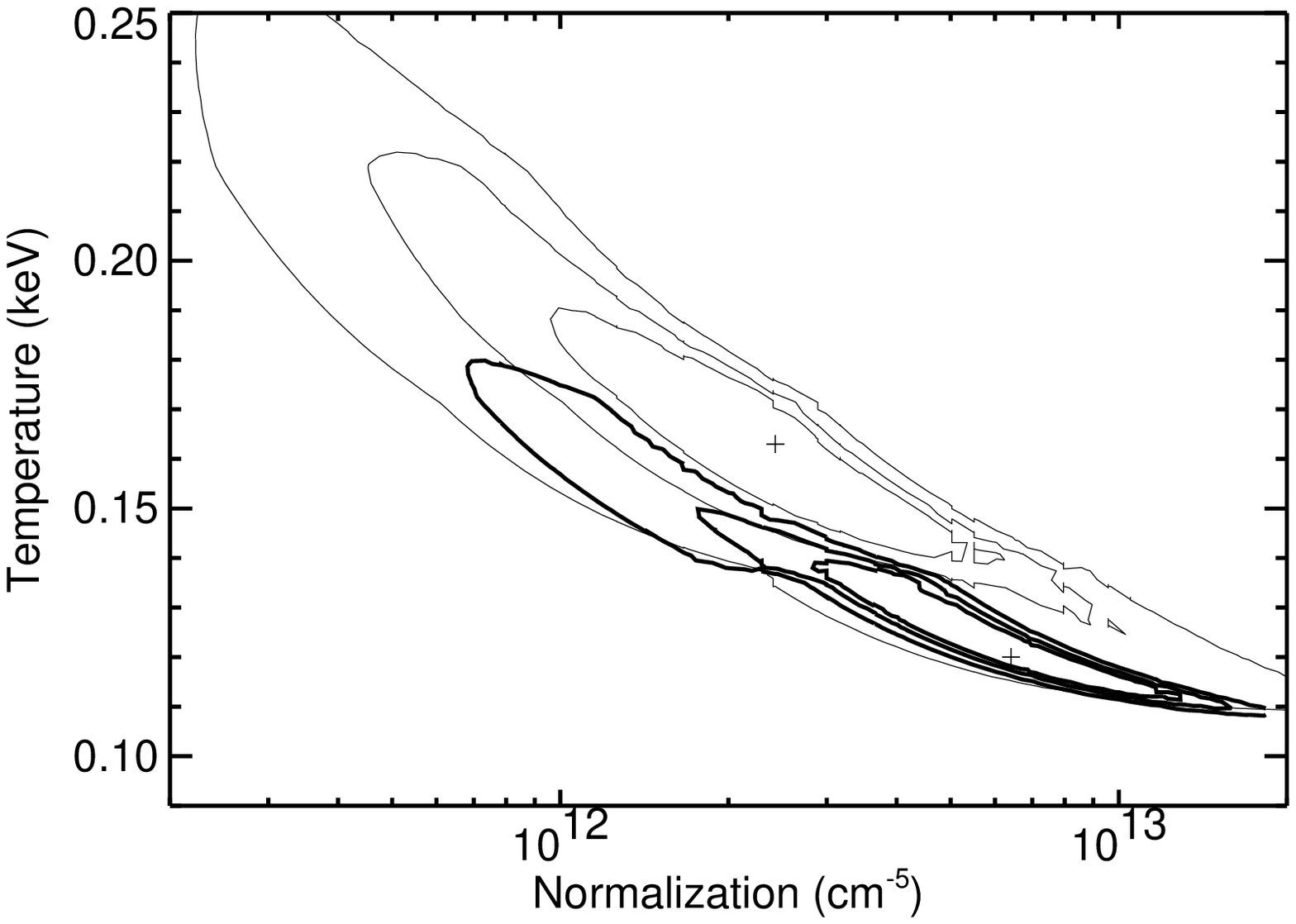}}
   \caption{68\%, 90\% and 99\% confidence level contours for the plasma temperature and emission measure of the thermal component used in the fit of the North Spur XMM-Newton EPIC ({\em thin lines}) and Chandra ACIS-S ({\em thick lines}) spectra.}
   \label{ktnorm}
   \end{figure}
Fig. \ref{ktnorm} shows the $\chi^2$ contour levels in the temperature-emission measure plane corresponding to confidence levels of 68\%, 90\% and 99\% on parameter uncertainties.
   \begin{figure}[tbp!]
   \centerline{\includegraphics[width=8.0cm]{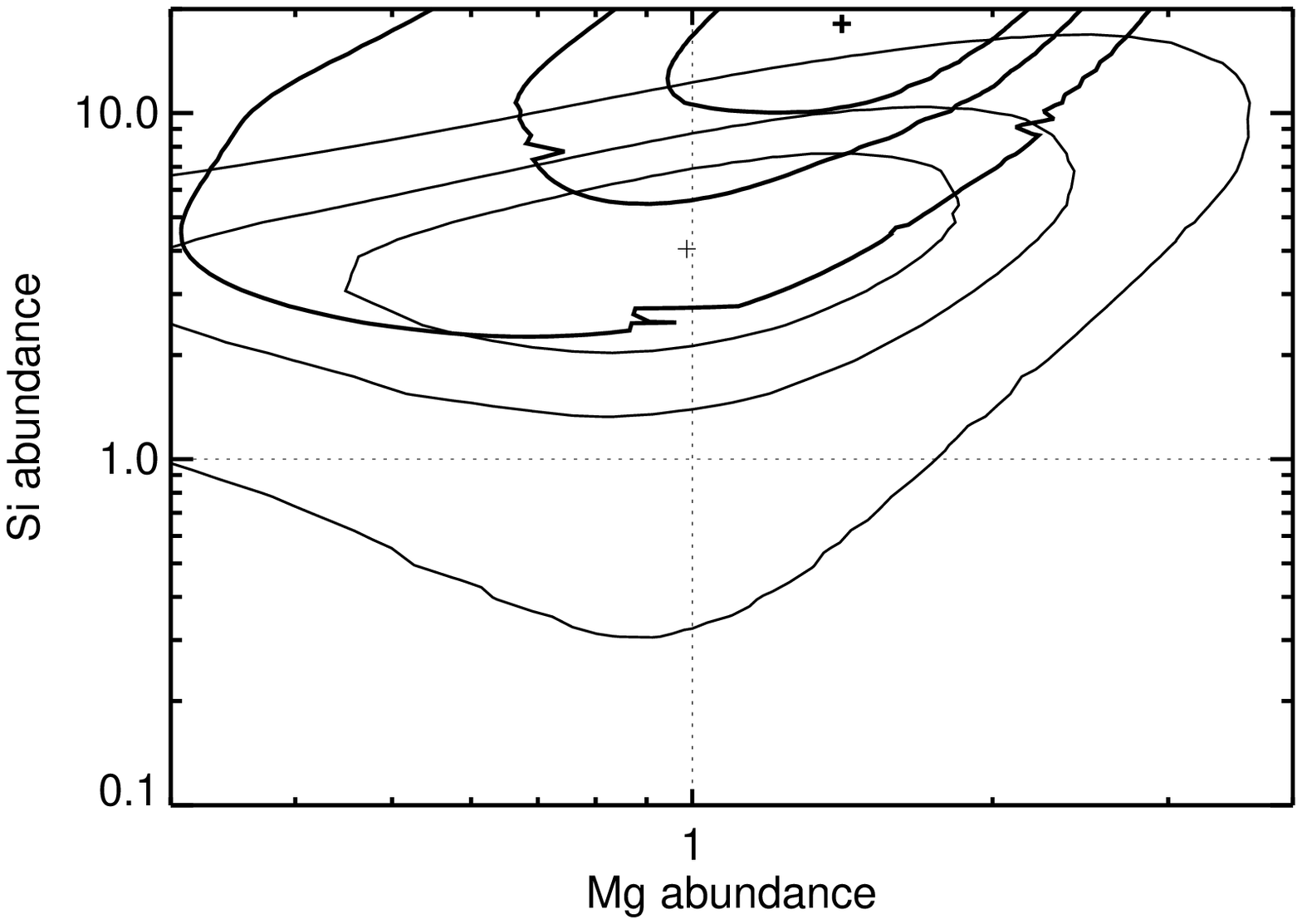}}
   \caption{Same as Fig. \protect\ref{ktnorm} but for Mg and Si abundances relative to solar.}
   \label{mgsi}
   \end{figure}
Fig. \ref{mgsi} shows the same contours levels of Fig. \ref{ktnorm},
but for the Mg and Si abundances when these parameters are left free to
vary. The $\chi^2$ obtained in this case is better than the
case of abundances fixed to solar values (F-test probability of $2\times
10^{-3}$), but the solar values for both elemental abundances are still
within the 99\% confidence level.

\section{Discussion}

The best-fit Si abundances obtained from the fit is around 4 and its 90\%
confidence level range is between about 2 and 10 for the XMM-Newton EPIC
and above 5 for Chandra  ACIS-S. Even though the solar value cannot be
excluded at a 99\% confidence level for XMM-Newton, the fact that both Chandra
and XMM-Newton independently give a best-fit Si abundance above solar
can be considered a hint that this element is somewhat
overabundant with respect to solar. On the other hand, Mg abundance is
found to be compatible with the solar value.

The best-fit Si abundance is not compatible with the abundances
found in shocks of young SNR interacting with the circumstellar material
(see, e.g., \citealt{phs04}). Instead, it may indicate that the North
Spur is an ejecta knot. The fact that the Mg abundance ranges between
0.3 and 2 with a best-fit of $\sim 1$ may imply that we are seeing at a
Si-rich knot like the one observed in Cas A (\citealt{lh03}) and in other
remnant. However, given the lack of other emission line in the spectrum
and the overall weakness of the thermal component, a full comparison
with the abundance patterns observed in ejecta of other young SNRs
is not possible, so the interpretation of North Spur as
a Si-rich ejecta knot needs more data to be confirmed.

Assuming a reasonable line of sight of the emitting plasma in the North
Spur (about 1 pc), a density of the knot of $3.8 (1.6-6) \times
d_5^{-1/2}$ cm$^{-3}$ were derived (where $d_5$ is the distance in units of 5 kpc,
\citealt{bs81}), and a total mass of the thermal plasma of $0.7 (0.3-1)
\times d_5^{-5/2}$ M$_\odot$. Both density and mass are consistent with
an ejecta of a Type II SN. However, the measured temperature (between
0.12 and 0.2 keV, considering both the EPIC and ACIS measurements)
is lower then typical ejecta knot temperature measured in young SNRs
(e.g. \citealt{lh03} for Cas A).

This discrepancy can be eliminated if we assume that the ejecta we are
observing have not been heated by the passage of the reverse shock, but by
the supersonic shock of the expanding pulsar wind nebula. The expansion
of a PWN in the freely expanding ejecta of its SNR has been studied
by several authors (\citealt{rc84}, \citealt{vag01}, \citealt{che04}).
The X-ray emission associated to the PWN supersonic shock into the ejecta
have been observed up to now only in 3C58 (\citealt{bwm01}), while for
the Crab and 0540-69 has been observed in the optical (\citealt{sh97},
\citealt{kmw89}). There are no other established observational signature
of PWN expanding into ejecta (see \citealt{che04} for a review). In
3C58, \citet{bwm01} measured an X-ray temperature of 0.2 keV, which
is remarkably similar to the one found here in the North Spur of
G21.5-0.9. However, the mass reported in the case of 3C58 was only 0.1
M$_\odot$ and \citet{che04} noted that these was higher then the predicted
swept-up mass of ejecta for the evolutionary stage of that remnant. If
we assume a similar evolutionary stage for G21.5-0.9, our mass estimate
in the North Spur is even more problematic for this scenario.

We note that the interpretation of the North Spur
in term of ejecta overrun by the PWN would imply a radius of the X-ray
plerion much larger then the radio plerion (about 120 arcsec vs. 50
arcsec of the radio plerion, \citealt{fhm88}), a feature which is not
expected and indeed not observed in any other PWN. Even if such a large
radius of the X-ray nebula compared to the radius of the radio nebula
was suggested elsewhere in literature for G21.5-0.9 (\citealt{shp01}),
it would be hard to explain it in terms of conventional PWN models.

Finally, we note that the differences of best-fit values between Chandra
and XMM in Table \ref{res} do not affect the conclusion of the paper
about the presence of an additional thermal component in the spectrum.
One of the reason of the differences may be the different extraction
radii used for Chandra and XMM (this fully explains the differences in
the fluxes, for instance). Further discussions about the nature of the
differences are outside the scope of this work.





%

%

\end{document}